\documentclass[superscriptaddress,showpacs,preprint]{revtex4}
\usepackage{graphicx}
\usepackage{epsfig}

\usepackage{epstopdf}
\usepackage{delarray}
\usepackage{braket}
\usepackage{amsmath, amssymb}

\begin{document}
\title{Retrit States Violating the KCBS Inequality and Necessary Conditions for Maximal Contextuality}

\author{F. Diker}
\affiliation{Faculty of Engineering and Natural Sciences, Sabanc{\i} University, Tuzla 34956, Istanbul, Turkey}
                
\author{Z. Gedik}
\affiliation{Faculty of Engineering and Natural Sciences, Sabanc{\i} University, Tuzla 34956, Istanbul, Turkey}

\date{\today}
\begin{abstract}
Since violations of inequalities implied by non-contextual and local hidden variable theories are observed, it is essential to determine the set of (non-)contextual states. Along this direction, one should determine the conditions under which quantum contextuality is observed. It is also important to determine how one can find maximally contextual qutrits. In this work, we revisit the Klyachko-Can-Binicio\u{g}lu-Shumovsky (KCBS) scenario where we observe a five-measurement state-dependent contextuality. We investigate possible symmetries of the KCBS pentagram, i.e., the conservation of the contextual characteristic of a qutrit-system. For this purpose, the KCBS operator including five cyclic measurements is rotated around the $Z$-axis. We then check a set of rotation angles to determine the contextuality and non-contextuality regions for the eigenstates of the spin-$1$ operator for an arbitrary rotation. We perform the same operation for the homogeneous linear combination of the eigenstates with spin values $+1$ and $-1$. More generally, we work on the real subgroup of the three dimensional Hilbert space to determine the set of (non-)contextual states under certain rotations in the physical Euclidean space $\mathbb{E}^3$. Finally, we show data on Euler rotation angles for which maximally contextual retrits (qutrits of the real Hilbert space) are found, and derive mathematical relations through data analysis between Euler angles and qutrit states parameterized with spherical coordinates.

\end{abstract}
\maketitle


\section{INTRODUCTION}

Since we are accustomed to observing classical correlations in daily life, it is always harder for us to understand quantum phenomena.
Two fundamental properties, quantum non-locality, and contextuality are particularly intriguing for the scientific community. The characteristics of Quantum Theory (QT) continue to attract attention and raise new questions that have yet to be answered. To say more clearly, new problems have arisen because new evidence of QT contradicts with classical predictions; namely, they cannot be well understood through the principle of locality. In light of these problems, non-contextual and local hidden variables (NLHV) have been examined; however, these have been refuted by theoretical and experimental works \cite{1,2,3,4,5,6}.

Two simple scenarios have been proposed to determine whether observed events can be explained by NLHV theory \cite{1,10}. In the first scenario, there is only one agent performing measurements on a physical system in a contextuality test. In the second scenario, two agents are measuring their respective qubit systems in a non-locality test. These tests concern the fundamental question of whether classically obtained inequalities can be violated by quantum correlated systems. The outcomes of these measurements yield results outside the classical range. As a result, quantum correlations are defined well enough; in other words, we do not need the hidden variable paradigm. As it turns out, quantum correlations are not only non-local but contextual as well. The contextuality of quantum correlations has been demonstrated experimentally \cite{11,12}.

Contextuality was tested in a qutrit-system \cite{10}, and it was shown that the classically obtained inequality for a qutrit state, which is called the Klyachko-Can-Binicio\u{g}lu-Shumovsky ($KCBS$) inequality, is violated. This was a remarkable result because it shows that qutrit states (a three-level quantum state) are intrinsically contextual, which is unique to QT.

The outline of this paper is as follows: In Section II, we provide a brief introduction to the KCBS inequality. In Section III, we investigate the KCBS inequality through the analysis of the rotation around the $Z$-axis in the physical Euclidean space $\mathbb{E}^3$. In sections IV, V, and VI we continue this rotational analysis of the KCBS inequality for the eigenstates of the spin-$1$ operator and the homogeneous linear combination of spin states with spin values $+1$ and $-1$. In Section VII, we investigate the subgroup of qutrit states with real probability amplitudes and classify them by (non-)contextuality under rotations with certain angles. In Section VIII, by making use of the collected data, we derive general formulas for spherical parameters of maximally contextual retrits depending on rotations in the physical space.

\section{The KCBS Inequality}

The simplest example of KS-like scenarios is KCBS, which includes five measurements performed on a qutrit state \cite{10}. A spin-1 system is a typical example of this. Instead of working with the spin projections, Klyachko et al. defined new observables,
\begin{equation}
A_i = 2{S_{i}}^2 - 1
\label{eq4}
\end{equation}  
where $S_{i}$ are spin-1 projection operators. 

In this way, one reduces the number of eigenvalues to two, $+1$ and $-1$. ${S_{i}}^2$ and ${S_{j}}^2$ (and hence $A_i$ and $A_j$) can be measured together if $i$ and $j$ correspond to orthogonal directions. It is possible to find five directions so that $A_i$ and $A_{i+1}$ can be measured simultaneously. If any quantum system exhibits a non-contextual feature, the following inequality must be satisfied:
\begin{equation}
\langle A_1 A_2 \rangle + \langle A_2 A_3 \rangle + \langle A_3 A_4 \rangle + \langle A_4 A_5 \rangle + \langle A_5 A_1 \rangle \geq -3.
\end{equation}
However, due to the contextual nature of quantum systems, the left hand side can be as small as $5-4\sqrt{5}\ \ (\simeq -3.94) $. In other words, depending on the quantum state measured we observe that the outcome exceeds the lower limit of the classical approach.

\section{Rotational invariance of the KCBS Inequality about symmetry axis}

The KCBS inequality is the most basic example for showing the contextual behaviour of a certain group of qutrit systems. It is quite simple, including only five measurements performed on a qutrit. To the best of our knowledge, it has not been investigated from the point of rotational symmetries in the physical Euclidean space $\mathbb{E}^3$, i.e., the conservation rules of contextual, or non-contextual properties of quantum systems. For this purpose, we will first look at the rotational invariance of the KCBS operator about the symmetry axis which is taken to be the $Z$-axis in our case. The usual rotation operator is
\begin{equation}
e^{-i \frac{S_z}{\hbar} \alpha}
\end{equation}  
which gives
\begin{equation}
    I - {S_z}^2 (1 - \cos{\alpha}) - i S_z \sin{\alpha}.
\end{equation}
One may obtain this equation after the power series expansion of the exponential operator. The matrix representation of the rotation operator is as follows:
\begin{equation}
\left(
\begin{array}{ccc}
 \cos (\alpha )-i \sin (\alpha ) & 0 & 0 \\
 0 & 1 & 0 \\
 0 & 0 & \cos (\alpha )+i \sin (\alpha ) \\
\end{array}
\right).
\end{equation} 
We rotate the KCBS operator around the $Z$-axis by making the following calculations:
\begin{equation}
e^{i \frac{S_z}{\hbar} \alpha}  \bigg[  A_1 A_2  +  A_2 A_3 +  A_3 A_4 +  A_4 A_5  +  A_5 A_1 \bigg]    e^{-i \frac{S_z}{\hbar} \alpha}
\end{equation}
which gives
\begin{equation}
\left(
\begin{array}{ccc}
 -5+2 \sqrt{5} & 0 & 0 \\
 0 & 5-4 \sqrt{5} & 0 \\
 0 & 0 & -5+2 \sqrt{5} \\
\end{array}
\right).
\end{equation}
This is the usual KCBS operator, which means that the outcomes of five cyclic measurements are invariant under the rotation around the symmetry axis of the pentagram ($Z$-axis). In other words, contextuality (or non-contextuality) is symmetrically conserved for spin-1 quantum states around the $Z$-axis.

\section{Contextuality Region for Neutrally Polarized Spin State}

We already know that the neutrally polarized spin state denoted by $\ket{0}$ results in the maximal violation of the KCBS inequality; therefore, it is contextual. However, the question is whether we can always observe contextuality for this state under any rotation in the physical space. To that end, the KCBS operator denoted by $S$  is rotated by the general rotation operator as follows:
\begin{equation}
    e^{i \frac{S_z}{\hbar} \alpha}  e^{i \frac{S_y}{\hbar} \beta}  e^{i \frac{S_z}{\hbar} \gamma} S e^{-i \frac{S_z}{\hbar} \gamma}  e^{-i \frac{S_y}{\hbar} \beta}  e^{-i \frac{S_z}{\hbar} \alpha}
\end{equation} 
where $\alpha$, $\beta$ and $\gamma$ are Euler angles. We obtain $S'$ which is the transformed KCBS operator. We then take the average
\begin{equation}
\bra{0} S' \ket{0} = (5 - 3 \sqrt{5}) \cos{2 \beta} - \sqrt{5}
\end{equation}
which depends only on $\beta$, i.e., the average value depends on the rotation around the $Y$-axis. The set of $\beta$ values for which 
\begin{equation}
\langle S' \rangle < -3
\label{eq10}
\end{equation}
gives us the contextuality region for the state $\ket{0}$. Thereby, the non-contextuality region can also be found. The inequality in Equation (\ref{eq10}) is satisfied for $-31.717^{\circ} < \beta < 31.717^{\circ}$ and $ 148.283^{\circ} < \beta < 211.717^{\circ}$. This is an important result since a maximally contextual state, which is a zero-spin state in our case, does not necessarily show contextuality for any five-measurement KCBS scenario.    

\section{Rotational Independence of Non-contextuality for Spin-$1$ States}

If we want to determine the certain symmetries for the KCBS pentagram, we must also check all possible rotations in the physical Euclidean space $\mathbb{E}^3$. We aim to find a group of five measurements performed on the eigenstates of the spin-1 operator, and hence we will determine if these states exhibit contextual behaviour. We already know that a neutrally polarized spin state gives a maximal violation of the KCBS inequality, therefore exhibiting contextual behaviour. We will check the other two. The general rotation operator we used in the previous section is 
\begin{equation}
    e^{-i \frac{S_z}{\hbar} \gamma}  e^{-i \frac{S_y}{\hbar} \beta}  e^{-i \frac{S_z}{\hbar} \alpha},
\end{equation}
\begin{figure}[t!]
\centering
  \includegraphics[width=0.7\linewidth]{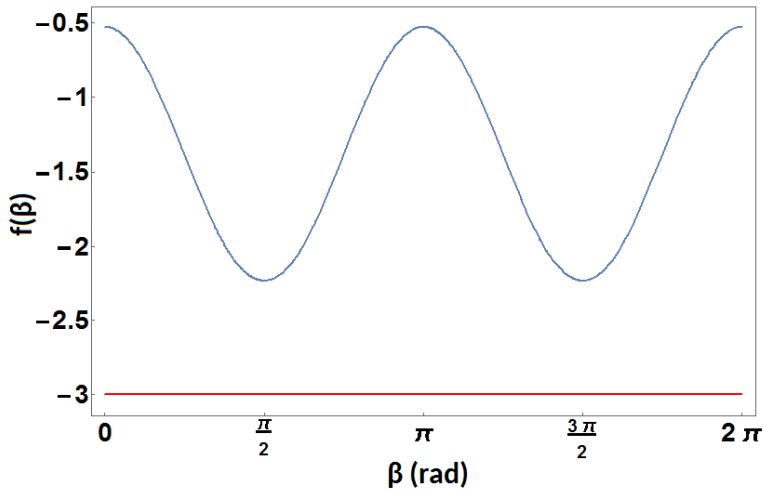}
\caption{The average value of the KCBS operator is a periodic function of $\alpha$ which repeats over intervals of $\pi$ radians. The function is denoted by $f(\beta)$, which does not exceed the classical limit, $-3$. This tells us that the spin states with spin values $+1$ and $-1$ are always independently non-contextual in the KCBS scenario.}
  \label{fig1}
\end{figure}
and the matrix representation of the general rotation operator is as follows:
\begin{equation}
    D( \alpha, \beta, \gamma) = \left(
\begin{array}{ccc}
 e^{-i \alpha -i \gamma } \cos ^2\left(\frac{\beta }{2}\right) & -\frac{e^{-i \gamma } \sin (\beta )}{\sqrt{2}} & e^{i \alpha -i \gamma } \sin ^2\left(\frac{\beta }{2}\right) \\
 \frac{e^{-i \alpha } \sin (\beta )}{\sqrt{2}} & \cos (\beta ) & -\frac{e^{i \alpha } \sin (\beta )}{\sqrt{2}} \\
 e^{i \gamma -i \alpha } \sin ^2\left(\frac{\beta }{2}\right) & \frac{e^{i \gamma } \sin (\beta )}{\sqrt{2}} & e^{i \alpha +i \gamma } \cos ^2\left(\frac{\beta }{2}\right) \\
\end{array}
\right).
\end{equation}
The KCBS operator is denoted by $S$, and the general rotation of $S$ is expressed as follows:
\begin{equation}
D^\dag (\alpha, \beta , \gamma)  S D( \alpha, \beta , \gamma) = S'(\alpha, \beta )
\end{equation} 
where $S'$ is a function of $\alpha$ and $\beta$. Recall that the $Z$-axis is taken as the symmetry axis of the KCBS inequality. The symmetry axis is also rotated, and the new symmetry axis is given by the following vector: $(\sin (\beta) \cos (\gamma), \sin (\beta) \sin (\gamma), \cos(\beta))$. The average of the new KCBS operator for the spin states is calculated as follows:
\begin{equation}
\bra{\pm 1} S' \ket{\pm 1} = \frac{1}{2} \left(\left(3 \sqrt{5}-5\right) \cos (2 \beta )+\sqrt{5}-5\right),
\end{equation} 
telling us that the only dependence is the rotation around the $Y$-axis. The plot for this result is illustrated in Figure 1.

\section{Contextuality Check for Homogeneous Linear Combination of Spin-1 States}

\begin{figure}[t!]
\centering
  \includegraphics[width=0.6\linewidth]{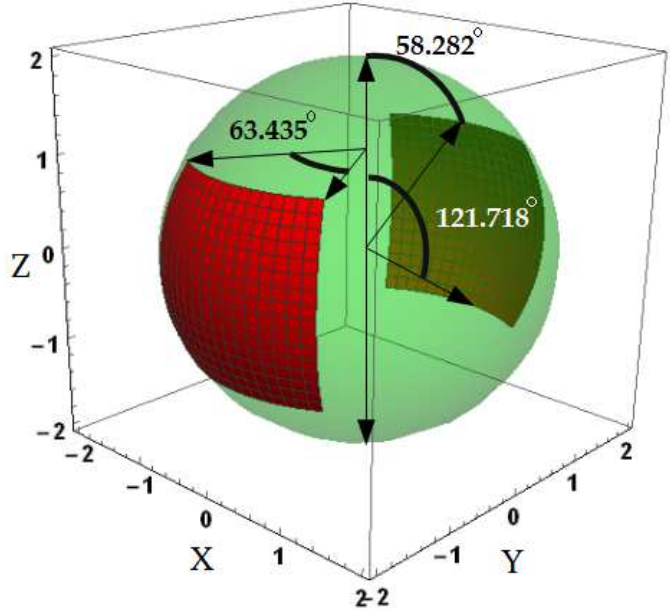}
\caption{Rotation of the KCBS operator in the physical space. Red surfaces on the sphere show contextuality regions for the homogeneous linear combination of $\ket{1}$ and $\ket{-1}$, which are the usual eigenstates of the spin-$1$ operator with eigenvalues $1$ and $-1$, respectively. The red parts of the sphere are spherically symmetric. The ranges of the rotation angles are shown in degree.}
  \label{fig2}
\end{figure}

We have seen that the spin states, $\ket{1}$ and $\ket{-1}$ are non-contextual independent of rotations in the physical space; in other words, they do not violate the KCBS inequality for any group of five measurements. We will now check if their homogeneous linear combination can violate the KCBS inequality. For this, we simply take the average of $S'(\alpha, \beta)$ as follows:
\begin{equation}
\bra{\psi} S'(\alpha,\beta) \ket{\psi} 
\end{equation}
where $\ket{\psi} = \frac{1}{\sqrt{2}} ( \ket{1} + \ket{-1})$. We get
\begin{equation}
\bra{\psi} S'(\alpha, \beta) \ket{\psi} =  \left(-5+2 \sqrt{5}\right) \cos ^2(\alpha )+\left(\left(-5+3 \sqrt{5}\right) \cos (2 \beta )-\sqrt{5}\right) \sin ^2(\alpha ),
\end{equation}
which yields values below the classical limit for the following sets of rotation angles: $58.282^{\circ} \le \beta \le 121.718^{\circ}$ and $58.282^{\circ} \le \alpha \le   121.718^{\circ}$, or $238.283^{\circ} \le \alpha \le   301.718^{\circ}$. In Figure 2, the ranges of rotation angles violating the KCBS inequality are shown on a sphere.  There are two minima for this function that is obtained for $\beta, \alpha = 90^{\circ}$ and $\alpha = 270^{\circ}$, $\beta = 90^{\circ}$. Both minimum values are $5-4 \sqrt{5}$, which means  the present quantum state maximally violates the KCBS inequality. The average of the spin-1 operator for the homogeneous linear combination of polarized states gives zero, as in the case of zero-spin state, and both $\ket{0}$ and $\ket{\psi}$ maximally violate the KCBS inequality. However, as mentioned earlier, the spin-$1$ states do not show contextual behaviour when measured separately in the KCBS scheme.

\section{Classification of Retrit States into Contextual and Non-Contextual Groups}

We investigate the real Hilbert space for qutrit states to check their contextuality for an arbitrary rotation of the KCBS operator in the physical Euclidean space $\mathbb{E}^3$. The subgroup we work on spans all possible linear combinations of $\ket{0}$, $\ket{1}$ and $\ket{-1}$ with real probability amplitudes. The general expression of our state is
\begin{equation}
\ket{\psi} = a \ket{1} + b \ket{0} + c\ket{-1}
\end{equation}
where $a, b, c \in \mathbb{R}$. One may use spherical coordinates to express general retrit state; $\ket{\psi} = (\sin \theta  \cos \phi , \sin \theta  \sin \phi , \cos \theta )$ where $0 \leq \theta < \pi$ and $0 \leq \phi < 2\pi$. We perform the same operation as we did before and rotate the KCBS operator. We then take the average as follows:
\begin{equation}
\bra{\psi} D^\dag ( \alpha, \beta, \gamma)  S D( \alpha, \beta, \gamma) \ket{\psi}
\end{equation}
which gives a function of $\alpha$, $\beta$, $\theta$ and $\phi$,
\begin{equation}
\begin{split}
 f(\theta, \phi, \beta, \alpha) = \frac{1}{4} \bigg( 2 \left(\left(3 \sqrt{5}-5\right) \cos (2 \beta )+\sqrt{5}-5\right) \cos ^2(\theta)     
\\+\left(3 \sqrt{5}-5\right) \sin ^2(\theta) \left(\cos (2 \beta ) (3 \cos (2 \phi)-1)-2 \sqrt{2} \sin (2 \beta ) \cos (\alpha ) \sin (2 \phi)\right)
\\+4 \left(3 \sqrt{5}-5\right) \sin (\beta ) \sin (2 \theta) \left(\sin (\beta ) \cos (2 \alpha ) \cos (\phi)+\sqrt{2} \cos (\beta ) \cos (\alpha )
\sin (\phi) \right)
\\+\sin ^2(\theta) \left(\left(3 \sqrt{5}-5\right) \cos (2 \phi)-\sqrt{5}-5\right) \bigg).
\end{split}
\label{eq1}
\end{equation}
The analytic solution of this function with four variables is challenging; however, one may look into retrit states rotated around the $Y$- and $Z$-axis with specific angles. By assigning values to $\beta$ and $\alpha$, one may reduce Equation \ref{eq1} to a two-variable function. From the experimentalist point of view, this will provide a guideline for possible experiments that can be realized in the future. 

\begin{figure}[t!]
\centering
  \includegraphics[width=0.7\linewidth]{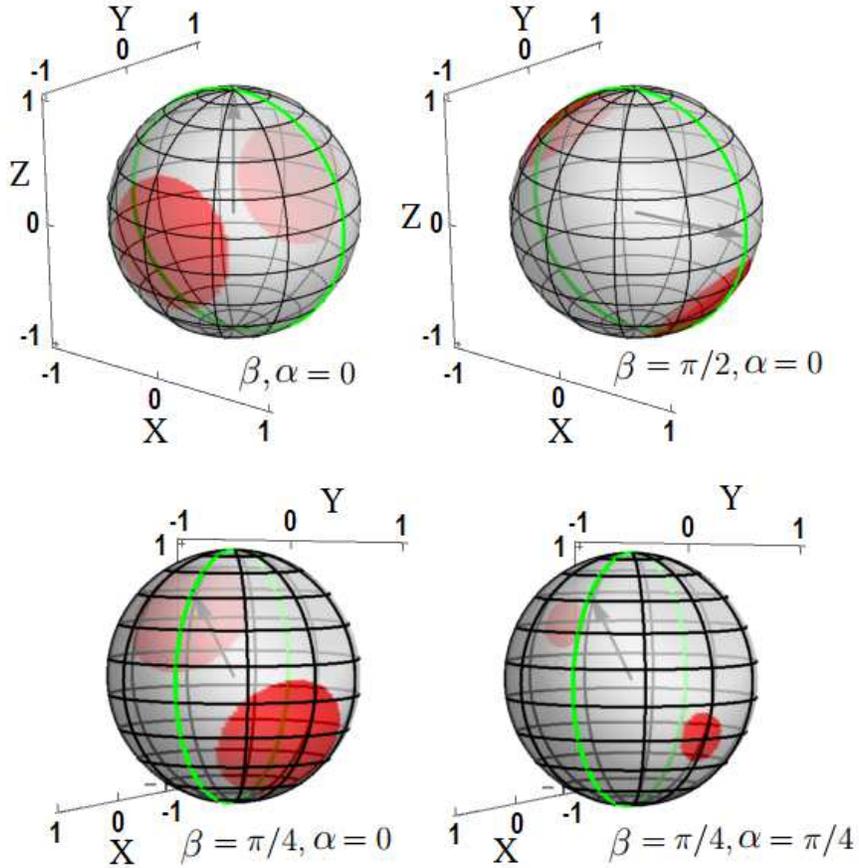}
\caption{For the general retrit state parameterized with spherical coordinates, contextuality regions are illustrated on four spheres with different rotation angles. The regions shown in red include the set of states violating the KCBS inequality. The light grey parts are for the non-contextuality regions, i.e., measurements yield inside the classical range. The thick black arrow is the symmetry axis of the KCBS pentagram, and the great circle connecting poles is composed of two bows at $\phi=0$ and $\phi=\pi$. Here, it is explicitly shown that the red contextuality surfaces rotate and shrink as the KCBS operator is rotated.  }
  \label{fig3}
\end{figure}

The first example is when there is no rotation; in other words, $\beta, \alpha = 0$ which results in
\begin{equation} \label{sp1}
\begin{split}
 f(\theta, \phi, 0, 0) = \frac{1}{4} \bigg( \sin ^2(\theta) \left(\left(3 \sqrt{5}-5\right) \cos (2 \phi)-\sqrt{5}-5\right)
\\+\left(3 \sqrt{5}-5\right) \sin ^2(\theta) (3 \cos (2 \phi)-1)+2 \left(4 \sqrt{5}-10\right) \cos ^2(\theta) \bigg).
\end{split}
\end{equation}
Through straightforward calculations one may find $f(\theta, \phi, \beta, \alpha)$ for different rotation angles. The latter examples we check are obtained for rotations around only the $Y$-axis, or both $Y$- and $Z$-axis, which are provided in the following:
\begin{equation}\label{sp2}
\begin{split}
f(\theta, \phi, \pi/2, 0) = \frac{1}{4} \bigg(\sin ^2(\theta) \left(\left(3 \sqrt{5}-5\right) \cos (2 \phi)-\sqrt{5}-5\right)
\\+\left(3 \sqrt{5}-5\right) \sin ^2(\theta) (1-3 \cos (2 \phi))
\\+4 \left(3 \sqrt{5}-5\right) \sin (2 \theta) \cos (\phi)-4 \sqrt{5} \cos ^2(\theta)\bigg),
\end{split}
\end{equation}
\begin{equation}\label{sp3}
\begin{split}
f(\theta, \phi, \pi/4, 0) = \frac{1}{8} \bigg(2 \left(3 \sqrt{5}-5\right) \sin ^2(\theta) \left(\cos (2 \phi)-2 \sqrt{2} \sin (2 \phi)\right)
\\+4 \left(3 \sqrt{5}-5\right) \sin (2 \theta) \left(\sqrt{2} \sin (\phi)+\cos (\phi)\right)+\left(3 \sqrt{5}-5\right) \cos (2 \theta)+\sqrt{5}-15\bigg),
\end{split}
\end{equation}
and
\begin{equation}\label{sp4}
\begin{split}
f(\theta, \phi, \pi/4, \pi/4) = \frac{1}{8} \bigg(4 \left(3 \sqrt{5}-5\right) \sin (2 \theta) \sin (\phi)
\\+2 \left(3 \sqrt{5}-5\right) \sin ^2(\theta) (\cos (2 \phi)-2 \sin (2 \phi))
\\+\left(3 \sqrt{5}-5\right) \cos (2 \theta)+\sqrt{5}-15\bigg).
\end{split}
\end{equation}
In Figure \ref{fig3}, the region plots of equations \ref{sp1}-\ref{sp4} are illustrated on spherical surfaces, which is compatible with our parameterization. One may notice that the contextuality regions shown as red surfaces on the spheres rotate and  shrink when we rotate the KCBS operator around the $Y$- and $Z$-axis. We already know that the contextuality observed in the KCBS scenario is invariant under rotations around the $Z$-axis; however, rotations around the $Y$-axis together with rotations around the $Z$-axis make the contextual set of retrit states smaller or larger depending on rotation angles. This concludes that the contextual behaviour of qutrit states in the KCBS scenario is not always observed; in other words, one state may exhibit contextuality under certain circumstances whereas it may behave classically (non-contextually) under some other conditions. Furthermore, under rotations with certain angles no retrit state can violate the KCBS inequality, i.e., they yield results compatible with the non-contextual inequality. Here are two examples of the set of rotation angles where we do not observe any contexual retrit state:  (1) $52.1^{\circ} \le \beta \le 128^{\circ}$, $232.2^{\circ} \le \beta \le 308.1^{\circ}$ where $\alpha = 41.8^{\circ}$; (2) $32.9^{\circ} \le \beta \le 61.5^{\circ}$, $118.5^{\circ} \le \beta \le 147.2^{\circ}$, $213.0^{\circ} \le \beta \le 241.6^{\circ}$, and $298.7^{\circ} \le \beta \le 327.3^{\circ}$ where $\alpha = 104.5^{\circ}$.

\section{Mathematical Relations between Euler Angles and Maximally Contextual Retrit States}

In this section, we will establish mathematical relations between Euler rotation angles and maximally contextual retrit states through data analysis. Recall that the general rotation operator is
\begin{equation}
e^{-i \frac{S_z}{\hbar} \gamma} e^{-i \frac{S_y}{\hbar} \beta} e^{-i \frac{S_z}{\hbar} \alpha},
\end{equation}
where the last rotation around the $Z$-axis does not change the expectation value of the KCBS operator. So, in this case, finding the set of only $\beta$ and $\alpha$ values for which we observe a maximal violation of the KCBS inequality is sufficient. We parameterized retrit states with spherical coordinates as follows: $\ket{\psi} = (\sin \theta \cos \phi , \sin \theta \sin \phi , \cos \theta )$ where $0 \leq \theta < \pi$ and $0 \leq \phi < 2\pi$. We have four variables in total. Our main purpose here is to find the retrit states which exhibit maximal contextuality and their dependence on Euler rotation angles. So we need to find four relations for $\theta_{min}$ and $\phi_{min}$ which represent values for the maximal violation of the KCBS inequality. Note that the reason why we call parameters 'min' is because maximal violation means $f(\theta, \phi, \beta, \alpha)$ is at global minimum. One needs to solve $f(\theta, \phi, \beta, \alpha)$ for its minimum; however, an analytic solution to this is a challenging one even though we investigate only the real subgroup of qutrits. The number of variables is six when the problem is addressed in the most general form, making the solution even more challenging. For the most general case, an increase in the number of variables is because the general form of a qutrit has four parameters. Extra two parameters come from relative phases. In our case, as mentioned earlier, we have four variables in total, on which we acquired data giving the most contextual result. It is well known that
\begin{equation}
f(\theta_{min}, \phi_{min}, \beta_{min}, \alpha_{min}) = 5 - 4 \sqrt{5},
\label{min}
\end{equation}
which is the global minimum, i.e., the most contextual result. We collected some data on the set of $\theta_{min}$ and $\phi_{min}$ values corresponding to Euler rotation angles, $\beta_{min}$ and $\alpha_{min}$. The data is shown in Table 1. We use this data to find possible trendlines between Euler angles and retrits and to show them on simple two-dimensional plots.

First, we assigned some values to $\beta$ (rotation angle around the $Y$-axis) and found $\alpha$ (rotation angle around the $Z$-axis) values for which we find maximally contextual retrits. This is simply the collection of data satisfying Equation \ref{min}. Assigned values to $\beta$ are in the following:
\begin{equation}
\beta = 0, \frac{\pi}{16}, \frac{2\pi}{16},\frac{3\pi}{16},...,\frac{15\pi}{16},\pi
\end{equation}
for each of which one may find $\theta_{min}$, $\phi_{min}$ and $\alpha_{min}$ values to obtain the result in Equation \ref{min}. We have reached this result by finding the minimum of the function for each assigned values of $\beta$. The solution set of $\alpha_{min}$ contains only two values, 0 and $\pi$. This narrows down the solution set of the general case, and one may state that there is always a maximally contextual retrit for each $\beta$ value as long as $\alpha_{min} \in \{0, \pi \}$.

\begin{figure}[t!]
\centering
\includegraphics[width=0.8\linewidth]{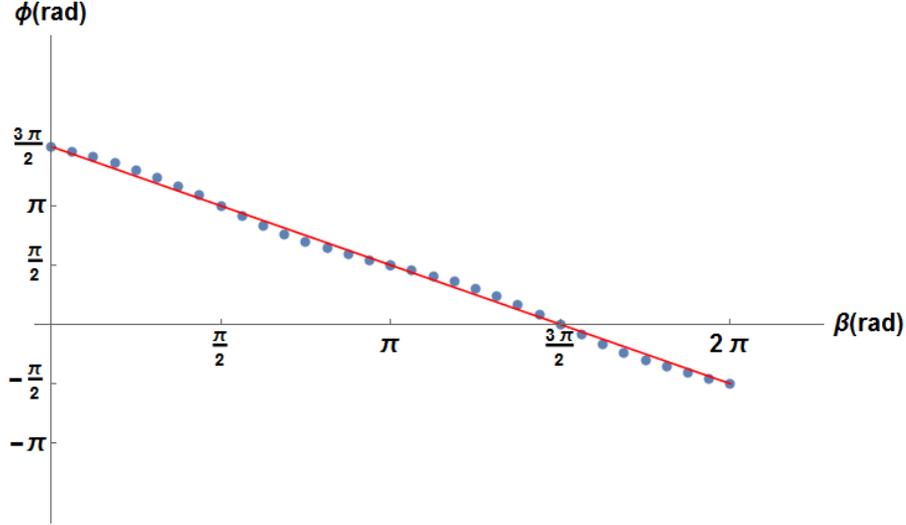}
\caption{The plot for $\phi_{min} (\beta)$ without the correction term. This is the trendline of the simple linear function for the data points taken from Table 1.}
\label{critical1}
\end{figure}

Next step is to find the mathematical relation between $\beta$ and $\phi$ ($\theta$) values. In Table 1, one may see $\phi$ and $\theta$ values for each $\beta$ ($\alpha$ is taken to be 0). In Figure \ref{critical1}, we illustrated data points and its trendline on $\beta$-$\phi$ graph, which is a straight line given by
\begin{equation}
\phi_{min} (\beta) = 4.71239\, - \beta.
\end{equation}
\begin{figure}[b!]
\centering
\includegraphics[width=0.7\linewidth]{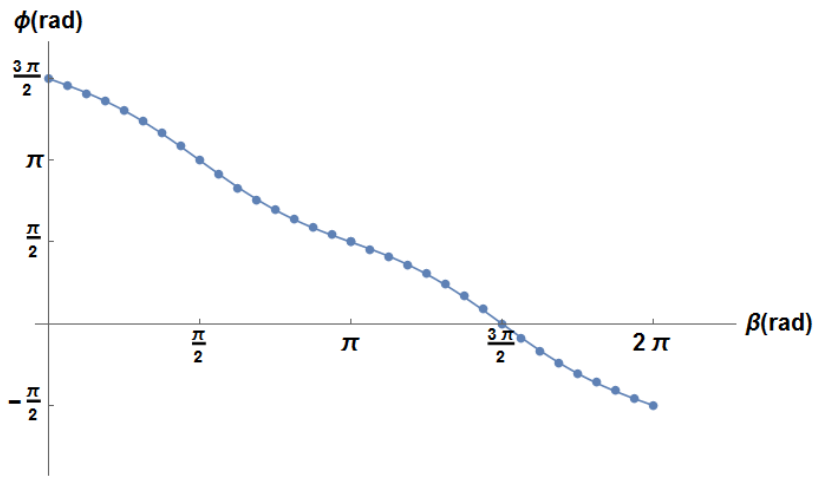}
\caption{The plot for $\phi_{min} (\beta)$ with the correction term. The curve fits better than the trendline of the linear function.}
\label{critical2}
\end{figure}
It is a straight line fitting almost perfectly; however, the collected data points follow a curve fluctuating slightly around the linear curve. Adding a trigonometric term with a small coefficient is a proper way because there is a slight fluctuation. A correction to the linear function gives us a curve fitting better, and we express $\phi_{min}(\beta)$ as:
\begin{equation}
4.71239 - \beta + (\frac{10}{59}) \sin (2 \beta).
\end{equation}
The plot for the corrected function is shown in Figure \ref{critical2}. Our next goal is to find how $\theta_{min}$ changes through the whole rotation around the $Y$-axis. Recall that $\alpha_{min}$ is taken to be $0$. The collected data points and the corresponding curve are illustrated in Figure \ref{critical3}. One may easily see that this is a usual trigonometric function with intervals of $2 \pi$. With some corrections to the sine function, we may obtain the following:
\begin{equation}
1.57 - 0.77 \sin (\beta).
\end{equation}
We have so far obtained the mathematical relations through data analysis, providing the set of Euler rotation angles for maximally contextual retrits in the KCBS scenario. Equations 5 and 6 provide the guideline for experimentalists to observe maximal quantum contextuality. As mentioned earlier, the rotation angle around the $Z$-axis, $\alpha_{min}$, can have only two values, $0$ and $\pi$ for maximal contextuality. For the relations obtained above, we looked into the case where $\alpha_{min}=0$. For $\alpha_{min}=\pi$, one may obtain similar relations due to symmetry.
\begin{figure}[t!]
\centering
\includegraphics[width=0.7\linewidth]{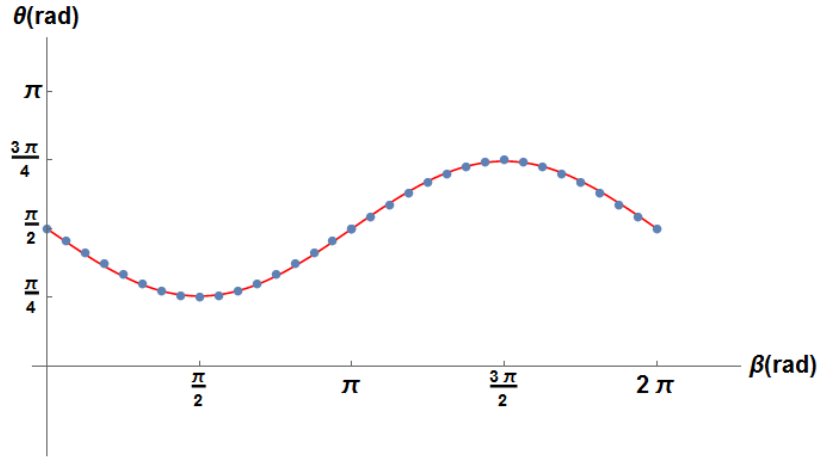}
\caption{The plot for $\theta_{min} (\beta)$. This is a simple plot of the sine function with a coefficient and a scalar addition.}
\label{critical3}
\end{figure}

\section{Conclusion}

The KCBS inequality is the most basic example of KS-like inequalities, and its significance comes from its simplicity. Therefore, the rotational-symmetry analysis of the KCBS operator is of the essence. Along this direction, we have shown that it is symmetrical about the $Z$-axis as expected. We have determined the contextuality region for $\ket{0}$ state, which gives a maximal violation for the usual KCBS operator. We have also checked the other eigenstates of the spin-$1$ operator, and shown that they are always non-contextual independent of rotation. We have seen that their homogeneous linear combination violates the KCBS inequality.

We have then investigated the specific set of qutrits with real probability amplitudes, which are called retrits, and provided some examples of physical rotations to determine the set of contextual retrits through graph analysis. We have explicitly shown the (non-)contextuality regions on spheres. Further, we have found the set of rotation angles for which no retrit state can violate the KCBS inequality, i.e., we have determined the set of KCBS measurements yielding results inside the classical range.

Finally, we have collected data for specific Euler angles for the KCBS scenario, giving maximally contextual retrits. We have shown data points on plots and found general formulas by fitting the corresponding curve on each plot. This has given us the general relation which must be satisfied for a retrit state to be maximally contextual. This suggests further research and paves the way for the classification of quantum systems of all kinds concerning their (non-)contextuality, and the degree of contextuality. Our results can be easily verified by experimentalists because the technique used in this work can be realized with current technology.


\appendix
\section{EULER ANGLES AND SPHERICAL PARAMETERS OF \\ 
RETRITS}
As mentioned earlier, finding all retrit states exhibiting maximal contextuality is a hard problem. Therefore, we have obtained data for different examples of maximally contextual retrits. By using the data, we have found curves fitting well enough for a possible solution to the problem. Here, we provide the data on Euler rotation angles and spherical parameters of retrits for maximal contextuality in the KCBS scenario. Note that $\alpha_{min}$ is taken to be zero.

\begin{table}[h]
\begin{tabular}{ |p{3cm}||p{3cm}|p{3cm}|p{3cm}|  }
 \hline
 \multicolumn{3}{|c|}{Table 1} \\
 \hline
 \multicolumn{1}{|c||}{Euler Angle} & \multicolumn{2}{|c|}{Spherical Parameters of Retrits} \\
 \hline
 $\beta_{min}\ \ (rad)$& $\theta_{min}\ \  (rad)$& $\phi_{min}\ \  (rad)$\\

\hline
0             & 1.57080 & 4.71239   \\
\hline
$\pi/16$ & 1.43240 & 4.57265   \\
\hline
$2 \pi/16$   & 1.29678 & 4.42746 \\
\hline
$3 \pi/16$  &  1.16707 & 4.27100  \\
\hline
$4 \pi/16$   &1.04720  & 4.09691  \\
\hline
$5 \pi/16$    & 0.94229 & 3.89869   \\
\hline
$6 \pi/16$    & 0.85888 & 3.67149 \\
\hline
$7 \pi/16$   & 0.80443 & 3.41581   \\
\hline
$8 \pi/16$   &0.78540  & 3.14159   \\
\hline
$9 \pi/16$ &0.80443  & 2.86737   \\
\hline
$10 \pi/16$  & 0.85888 & 2.61169 \\
\hline
$11 \pi/16$ & 0.94229 & 2.38449   \\
\hline
$12 \pi/16$  &  1.04720& 2.18628 \\
\hline
$13 \pi/16$   & 1.16707 & 2.01218  \\
\hline
$14 \pi/16$   &1.29678  & 1.85572    \\
\hline
$15 \pi/16$  & 1.43240 & 1.71053    \\
\hline
$16 \pi/16$ & 1.57080 & 1.57080 \\
\hline
$17 \pi/16$   &1.70919  & 1.43106  \\
\hline
$18 \pi/16$   &1.84481  & 1.28587  \\
\hline
$19 \pi/16$  & 1.97452 & 1.12941   \\
\hline
$20 \pi/16$  &2.09439  & 0.95532  \\
\hline
$21 \pi/16$  & 2.19930 & 0.75710  \\
\hline
$22 \pi/16$  & 2.28271 & 0.52990   \\
\hline
$23 \pi/16$   & 2.33716 & 0.27422  \\
\hline
$24 \pi/16$  &2.35619 & 0              \\
\hline

\end{tabular}
\end{table}

\begin{table}[t!]
\begin{tabular}{ |p{3cm}||p{3cm}|p{3cm}|p{3cm}|  }

 \hline
 \multicolumn{1}{|c||}{Euler Angle} & \multicolumn{2}{|c|}{Spherical Parameters of Retrits} \\
 \hline
 $\beta_{min}\ \ (rad)$ & $\theta_{min}\ \  (rad)$& $\phi_{min}\ \  (rad)$\\

\hline
$25 \pi/16$ & 2.33716 & -0.27422 \\
\hline
$26 \pi/16$ & 2.28271 & -0.52990 \\
\hline
$27 \pi/16$  & 2.19930 & -0.75710  \\
\hline
$28 \pi/16$  &2.09439 & -0.95532 \\
\hline
$29 \pi/16$   & 1.97452 & -1.12941   \\
\hline
$30 \pi/16$  & 1.84481 & -1.28587  \\
\hline
$31 \pi/16$   & 1.70919 & -1.43106 \\
\hline
$32 \pi/16$   & 1.57080 & -1.57080 \\
\hline

\end{tabular}
\end{table}


\end{document}